\documentclass[12pt]{article}

\usepackage{amsmath}
\usepackage{fullpage}
\usepackage{amssymb}
\usepackage{amsthm}
\usepackage{hyperref}
\usepackage{setspace}
\usepackage{graphicx}
\usepackage{bm}
\usepackage{physics}
\usepackage{authblk}

\begin{document}

 \title{The chemical birth-death process with Gillespie noise}
 \author{John J. Vastola}
\affil{Department of Physics and Astronomy, Vanderbilt University, \\ Nashville, Tennessee}

 \maketitle

\begin{abstract}
A nontrivial technical issue has long plagued the literature on stochastic path integrals: it is not clear which definition is correct in the case of multiplicative/state-dependent noise. One reason for this is the unavailability of exactly solvable toy problems with state-dependent noise, that could in principle be used to compare the correctness of different approaches. In this paper, we provide an exact path integral calculation of the transition probability corresponding to a one-dimensional system with state-dependent noise. In particular, we solve the chemical birth-death process with Gillespie noise (the canonical continuous approximation to the discrete birth-death process often used as a toy model in chemical kinetics) using a Martin-Siggia-Rose-Janssen-De Dominicis (MSRJD) path integral. We verify that our result is correct by solving the Fokker-Planck equation via eigenfunction expansion. 
\end{abstract}
%%%%%%%%%%%%%%%%%%%%%%%%%%%%%%%%%%%%%%%%%%%

% Referees: Eichhorn? Seifert?

\section{Introduction}
\label{sec:intro}

Models of stochastic dynamics offer one promising route towards understanding non-equilibrium statistical mechanics \cite{crooks1999, derrida2007, kurchan2007, lubeck2004, chou2011}. Given stochastic models characterized by things like master equations and Fokker-Planck equations, many researchers have been able to construct useful notions of heat, work, entropy production, and other quantities familiar from equilibrium statistical mechanics and thermodynamics \cite{speck2007, sughiyama2013, jarzynski2017, saha2018, bo2019, neri2019, crosato2019}. Indeed, the project of stochastic thermodynamics is proceeding productively overall \cite{seifert2012, seifert2019}. 

Still, one occasionally runs into difficulties. For example, one technical tool in the non-equilibrium physicist's arsenal is the stochastic path integral (e.g. \cite{elgart2004, elgart2006, cohen2008, ghosal2016, budkov2016, greenman2016, greenman2017, greenman2018, caprini2019}), which can describe stochastic systems in a way that is on equal footing with stochastic differential equation (SDE) and Fokker-Planck descriptions \cite{graham1977, kleinert2009, hertz2016, weber2017, vastola2019}, and that complements those descriptions in theoretical analyses of a given system. Interestingly, there is a lack of consensus in the literature regarding the correct path integral treatment of Langevin equations with state-dependent/multiplicative noise \cite{arnold2000, arenas2010, tang_summing_2014, cugliandolo2017, cugliandolo2019}.  Rather than being merely an issue of preference, the presence or absence of different terms in the action of the path integral can lead to qualitatively different predictions (e.g. for optimal transition paths) for moderate amounts of noise\footnote{See Sec. 6.1 of Wang \cite{wang2015}, for example, where an optimal path missing a saddle point is attributed to a Jacobian-derived term in the action. This term is not present in some other path integral actions \cite{weber2017, vastola2019}.}. 

One reason it is hard to compare the correctness of different path integral descriptions is that path integrals are almost always computed approximately---it is \textit{hard} to find nontrivial problems which are exactly solvable. In the case of stochastic path integrals, it is apparently hard to find nontrivial and exactly solvable toy problems with state-dependent noise. 

In this paper, we present an exact path integral solution for a one-dimensional stochastic system described by an SDE with state-dependent noise. In particular, we solve for the transition probability $P(x, t; x_0, t_0)$ associated with the Ito-interpreted chemical Langevin equation (CLE)
\begin{equation} \label{eq:CLE}
\dot{x} = k - \gamma x + \sqrt{k + \gamma x} \ \eta(t)
\end{equation}
where $\eta(t)$ is a Gaussian white noise term, and where $x \in [-k/\gamma, \infty)$ (since there is nonzero probability of the system being pushed in the negative direction while the magnitude of the noise function is nonzero). This SDE is the canonical continuous approximation \cite{gillespie2000, gillespie2013} to the chemical birth-death process often used as a toy model in studies of chemical kinetics and gene regulation \cite{fox2017, bressloff2017, munsky2018}. 

Because Gillespie was the first to rigorously justify---without invoking the thermodynamic limit---the noise term in Eq. \ref{eq:CLE}, we will call this system the chemical birth-death process with Gillespie noise. It is closely related to an even simpler approximation (which is only valid when the system is sufficiently close to its steady state value $x_{ss} = k/\gamma$) of the same system, the chemical birth-death process with additive noise \cite{vastolaADD2019}. 

We will exactly solve for the transition probability $P(x, t; x_0, t_0)$ using the Martin-Siggia-Rose-Janssen-De Dominicis (MSRJD) path integral described in \cite{vastola2019}. In order to verify the correctness of our path integral calculation, we will first compute the transition probability corresponding to Eq. \ref{eq:CLE} by the method of eigenfunction expansion \cite{risken1996}. We do not undertake the project of comparing and contrasting every possible approach to stochastic path integrals; we merely show that ours reproduces the correct result in this specific case, lending credence to the idea that it is correct. 

The paper proceeds as follows. In Sec. \ref{sec:eigexpansion}, we derive the ground truth for $P(x, t; x_0, t_0)$ using the textbook method of eigenfunction expansion. In Sec. \ref{sec:pathintcalc}, we use the MSRJD path integral to derive the same result via a lengthy calculation. Finally, in Sec. \ref{sec:discussion} we discuss some consequences of this calculation for users of stochastic path integral approaches.

%======================================

\section{Eigenfunction expansion solution}
\label{sec:eigexpansion}

In this section, we compute the transition probability $P(x, t; x_0, t_0)$ corresponding to Eq. \ref{eq:CLE} via the method of eigenfunction expansion \cite{risken1996}. This essentially amounts to solving the Fokker-Planck equation, which describes how the system's probability density function evolves in time, using separation of variables. 

Because it will appear in many of the expressions to follow, define $\mu := k/\gamma$. The parameter $\mu$ corresponds to the steady state value of $x$ (to show this, set the right-hand side of Eq. \ref{eq:CLE} to zero), and we will see that the steady state probability distribution $P_{ss}$ only depends on $\mu$, and not on $k$ or $\gamma$ separately. It will also appear in the bounds of integrals, since $x \in [-\mu, \infty)$. 

Let $P(x, t)$ denote the probability that the system corresponding to Eq. \ref{eq:CLE} is in state $x \in [-\mu, \infty)$ at time $t \geq t_0$, subject to the initial condition that $P(x, t_0) = P_0(x)$ for some initial distribution $P_0(x)$. The Fokker-Planck equation for $P(x, t)$ reads
\begin{equation} \label{eq:TDFP}
\begin{split}
\frac{\partial P(x,t)}{\partial t} &= - \frac{\partial}{\partial x} \left[ (k - \gamma x) P(x,t) \right] + \frac{1}{2} \frac{\partial^2}{\partial x^2} \left[ (k + \gamma x) P(x, t) \right] \\
&= \gamma P(x,t) - (k - \gamma - \gamma x) \frac{\partial P(x,t)}{\partial x}  + \frac{(k + \gamma x)}{2} \frac{\partial^2 P(x,t)}{\partial x^2} \ .
\end{split}
\end{equation}
The boundary conditions are that
\begin{enumerate}
\item $\lim_{x \to \infty} P(x, t) = \lim_{x \to \infty} P'(x, t) = 0$,  and $P(x, t)$ dies off fast enough that the integral $\int P(x, t) dx$ converges for all $t$.
\item $P(-\mu, t) = 0$ for all $t \geq t_0$.
\end{enumerate} 
Together these conditions guarantee that, provided $P_0(x)$ was normalized, $P(x, t)$ remains normalized for all times $t$. Because we are specifically interested in the transition probability $P(x, t; x_0, t_0)$, we will assume the initial distribution was $P_0(x) = \delta(x - x_0)$ for some state $x_0 \in [-\mu, \infty)$. 

\subsection{Steady state Fokker-Planck solution}

As a starting point, we would like to find $P_{ss}(x)$, the steady state solution to the Fokker-Planck equation. Setting $\partial P/\partial t = 0$ in Eq. \ref{eq:TDFP}, we have
\begin{equation} \label{eq:ssFPd}
0 =  - \frac{\partial}{\partial x} \left[ (k - \gamma x) P_{ss}(x) \right] + \frac{1}{2} \frac{\partial^2}{\partial x^2} \left[ (k + \gamma x) P_{ss}(x) \right] \ .
\end{equation}
Integrate both sides (and note that the arbitrary constant that appears must be zero for both sides to vanish at infinity) to obtain the steady state Fokker-Planck equation
\begin{equation} \label{eq:ssFP}
0 =  (k - \gamma x) P_{ss}(x)  - \frac{1}{2} \frac{\partial}{\partial x} \left[ (k + \gamma x) P_{ss}(x) \right] \ .
\end{equation}
Solving this simple ODE and normalizing our result, we obtain
\begin{equation} \label{eq:pss}
P_{ss}(x) = \frac{2^{4 \mu}}{\Gamma(4 \mu)} \left( x + \mu \right)^{4 \mu - 1} \exp\left[ - 2 (x + \mu) \right] \ .
\end{equation}

\subsection{Eigenfunctions}

Applying the standard separation of variables ansatz $P(x, t) = P_E(x) T(t)$ to the Fokker-Planck equation (Eq. \ref{eq:TDFP}) yields the general solution
\begin{equation} \label{eq:FPgenslnE}
P(x, t) = \sum_{E} c_E P_E(x) e^{- E (t - t_0)}
\end{equation}
where the $c_E$ are chosen so that $P(x, t_0)$ equals some initial distribution $P_0(x)$, and where $P_E(x)$ satisfies the \textit{time-independent} Fokker-Planck equation
\begin{equation} \label{eq:TIFP}
-E P_E(x) = \gamma P_E(x) - (k - \gamma - \gamma x) \frac{\partial P_E(x)}{\partial x}  + \frac{(k + \gamma x)}{2} \frac{\partial^2 P_E(x)}{\partial x^2}\ .
\end{equation}
Assume that the solution to Eq. \ref{eq:TIFP} can be written $P_E(x) = Q_E(x) P_{ss}(x)$. Substituting this ansatz into Eq. \ref{eq:TIFP} and using Eq. \ref{eq:ssFP} to simplify the result yields the equation
\begin{equation}
0 = \frac{(k + \gamma x)}{2}  Q_E''(x) + (k - \gamma x) Q_E'(x) + E Q_E(x)    \ .
\end{equation}
for $Q_E(x)$. Define $w := 2 (x + \mu)$ and $\bar{E} := E/\gamma$. Our equation becomes
\begin{equation} \label{eq:qDE}
0 = w Q_E''(w) + (4 \mu - 1 + 1 - w) Q_E'(w) + \bar{E} Q_E(w) = 0 \ ,
\end{equation}
which is the Laguerre differential equation. By a standard argument, this equation has solutions which do not blow up at infinity only if $\bar{E}$ is a nonnegative integer $n$. Hence, the solutions to Eq. \ref{eq:qDE} are the generalized/associated Laguerre polynomials $L_n^{(\alpha)}(w)$ with $\alpha = 4\mu - 1$. 

For our later convenience, we will write our solutions as 
\begin{equation}
Q_n(x) = \sqrt{ \frac{\Gamma(4\mu) n!}{\Gamma(n + 4\mu)} } \ L_n^{(\alpha)}\left( 2 (x + \mu) \right) \ ,
\end{equation}
with the corresponding solutions to Eq. \ref{eq:TIFP} being
\begin{equation}
P_n(x) = Q_n(x) P_{ss}(x) = \sqrt{ \frac{\Gamma(4\mu) n!}{\Gamma(n + 4\mu)} } \ L_n^{(\alpha)}\left( 2 (x + \mu) \right) P_{ss}(x) \ .
\end{equation}

\subsection{The propagator}

The time-dependent solution to the Fokker-Planck equation is
\begin{equation}
P(x, t; x_0, t_0) = \sum_{n=0}^\infty c_n P_n(x) e^{- E_n (t - t_0)}
\end{equation}
with the constants $c_n$ chosen so that $P(x, t_0; x_0, t_0) = \delta(x - x_0)$. Invoke the orthogonality of the generalized Laguerre polynomials to write
\begin{equation}
\int_{0}^{\infty} L_n^{(\alpha)}(w) L_m^{(\alpha)}(w) \ w^{\alpha} e^{- w} \ dw = \frac{\Gamma(n + \alpha + 1)}{n!} \ \delta_{nm} \ .
\end{equation}
In terms of our functions, the orthogonality relationship reads
\begin{equation}
\int_{-\mu}^{\infty} Q_m(x) Q_n(x) P_{ss}(x) \ dx = \delta_{nm} \ .
\end{equation}
This relationship can be exploited to compute the coefficients $c_n$. Set $t = t_0$, multiply both sides by $Q_m(x)$, and integrate; we get
\begin{equation}
\begin{split}
\int_{-\mu}^{\infty} Q_m(x) \delta(x - x_0) \ dx &= \sum_{n=0}^\infty c_n \int_{-\mu}^{\infty} Q_m(x) Q_n(x) P_{ss}(x)\ dx \\
\implies \ \int_{-\mu}^{\infty} Q_m(x) \delta(x - x_0) \ dx &= \sum_{n=0}^\infty c_n \delta_{nm} \\
\implies \ \int_{-\mu}^{\infty} Q_m(x) \delta(x - x_0) \ dx &= c_m \\
\implies \ Q_m(x_0) &= c_m \ .
\end{split}
\end{equation}
Hence, the solution to Eq. \ref{eq:TDFP} is
\begin{equation} \label{eq:sumsln}
\begin{split}
P(x, t; x_0, t_0) &= \sum_{n=0}^\infty Q_n(x_0) P_n(x) e^{- E_n (t - t_0)} \\
&= 2^{4 \mu} \left( x + \mu \right)^{4 \mu - 1} e^{ - 2 (x + \mu) } \sum_{n=0}^\infty  \frac{n!}{\Gamma(n + 4\mu)} \ L_n^{(\alpha)}\left( 2 (x_0 + \mu) \right) L_n^{(\alpha)}\left( 2 (x + \mu) \right)   e^{- \gamma n T}
\end{split}
\end{equation}
where $T := t - t_0$. 

\subsection{Summing the propagator}

In order to evaluate the infinite sum in Eq. \ref{eq:sumsln}, we can use the Hardy-Hille formula \cite{srivastava1969, bateman1953}, which says that 
\begin{equation}
\sum_{n = 0}^{\infty} \frac{n!}{\Gamma(1 + \alpha + n)} L_n^{(\alpha)}(x) L_n^{(\alpha)}(y) \ t^n = \frac{1}{(x y t)^{\alpha/2} (1-t)} e^{- \frac{(x + y) t}{1 - t}} I_{\alpha} \left( \frac{2 \sqrt{x y t}}{1 - t}  \right)
\end{equation}
where $I$ is the modified Bessel function of the first kind. Directly applying this formula to Eq. \ref{eq:sumsln}, we have
\begin{equation} \label{eq:ptrans}
\begin{split}
P(x, t; x_0, t_0) &= 2^{4 \mu} \left( x + \mu \right)^{4 \mu - 1} e^{ - 2 (x + \mu) } \sum_{n=0}^\infty  \frac{n!}{\Gamma(n + 4\mu)} \ L_n^{(\alpha)}(w_0) L_n^{(\alpha)}(w)   \left(e^{- \gamma T} \right)^n\\
&= 2^{4 \mu} \left( x + \mu \right)^{4 \mu - 1} e^{ - 2 (x + \mu) }  \frac{e^{- \frac{(w_0 + w) e^{- \gamma T}}{1 - e^{- \gamma T}}} }{(w_0 w e^{- \gamma T})^{(4\mu - 1)/2} (1-e^{- \gamma T})} I_{4\mu - 1} \left( \frac{2 \sqrt{w_0 w e^{- \gamma T}}}{1 - e^{- \gamma T}}  \right)  
\end{split}
\end{equation}
where we remind the reader that $w := 2(x + \mu)$ and $w_0 := 2(x_0 + \mu)$. We should note that this whole derivation closely parallels the eigenfunction expansion derivation of $P(x, t; x_0, t_0)$ for the chemical birth-death process with additive noise in \cite{vastolaADD2019}.

\section{Exact MSRJD path integral calculation}
\label{sec:pathintcalc}

Stochastic path integral descriptions of continuous stochastic systems described by SDEs like Eq. \ref{eq:CLE} express the transition probability $P(x, t; x_0, t_0)$ in terms of an infinite number of integrals. In the case of the Onsager-Machlup path integral \cite{OM1953pt1, OM1953pt2, graham1977, hertz2016}, the transition probability is expressed in terms of many integrals over state space; in the case of the Martin-Siggia-Rose-Janssen-De Dominicis (MSRJD) path integral  \cite{msr1973, janssen1976, dd1976, ddpeliti1978, hertz2016}, it is expressed in terms of integrals over both state space and a set of auxiliary variables. By analogy with quantum mechanics, we will call these auxiliary quantities momentum variables, although they are also sometimes called response variables \cite{chow2015}. 

The difference between the Onsager-Machlup and MSRJD path integrals is essentially the same as the difference between the configuration space (integrate over $x$ paths) and phase space (integrate over $x$ and $p$ paths) path integrals in quantum mechanics \cite{shankar2011}. For a derivation of both kinds of path integrals, see \cite{vastola2019}. For discussion of how path integral descriptions of CLEs can be viewed as approximations to a path integral description of the chemical master equation, see \cite{vastolaCLE2019}.

It turns out that the Onsager-Machlup path integral for Eq. \ref{eq:CLE} is kind of daunting, because of integrals that go like $e^{-1/x}$. For this reason, we will try using the MSRJD path integral instead. Just as in quantum mechanics, some problems (like the hydrogen atom \cite{duru1982, kleinert2009}) are more suited to a `phase space' approach. Even though this approach doubles the number of integrals we must do, the hope is that we have $2N - 1$ \textit{tractable} integrals instead of $N - 1$ impossible ones. 

The MSRJD path integral corresponding to a general one-dimensional Ito-interpreted SDE 
\begin{equation} \label{eq:genSDE}
\dot{x} = f(x) + g(x) \ \eta(t) \ , \ x \in [a, \infty)
\end{equation}
for some real number $a$ is\footnote{In the language of \cite{vastola2019}, Eq. \ref{eq:genSDE} describes the dynamics of an Ito-interpreted concentration-type variable (i.e. a variable whose domain is half-infinite).} \cite{vastola2019}
\begin{equation} \label{eq:MSRJD}
\begin{split}
P &= \lim_{N \to \infty} \int \frac{dp_N}{2\pi}\prod_{j = 1}^{N-1} \ \frac{dx_j dp_j}{2\pi} \ \exp{- \sum_{j = 1}^N \left[ i p_j \left( \frac{x_j - x_{j-1}}{\Delta t} - f(x_{j-1}) \right) + \frac{1}{2} {p_j}^2 g(x_{j-1})^2 \right] \Delta t}  \\
&= \int \mathcal{D}[x(t)] \mathcal{D}[p(t)] \ \exp\left\{- S[x, p] \right\} 
\end{split}
\end{equation}
where $P$ is shorthand for $P(x, t; x_0, t_0)$, $\Delta t := (t-t_0)/N$, $x_N := x$, and the second line is the `schematic' representation of the MSRJD path integral. We will be working with the discretized path integral directly, and so will avoid subtleties associated with taking the continuum limit. 

In our specific case, we have
\begin{equation} \label{eq:MSRJD_gill}
P = \lim_{N \to \infty} \int \frac{dp_N}{2\pi}\prod_{j = 1}^{N-1} \ \frac{dx_j dp_j}{2\pi} \ \exp{- \sum_{j = 1}^N \left[ i p_j \left( \frac{x_j - x_{j-1}}{\Delta t} - k + \gamma x_{j-1} \right) + \frac{1}{2} {p_j}^2 (k + \gamma x_{j-1}) \right] \Delta t}  \ .
\end{equation}
The integrations over the momentum variables $p_j$ are from $-\infty$ to $\infty$, and the integrations over the concentration variables $x_j$ are from $-\mu$ to $\infty$. The first thing we will do is to change to simpler variables. Define $y_j = \mu + x_j$ for $j = 0, 1, ..., N$. The Jacobian is trivial, so we now have
\begin{equation}
P = \lim_{N \to \infty} \int \frac{dp_N}{2\pi}\prod_{j = 1}^{N-1} \ \frac{dy_j dp_j}{2\pi} \ \exp{- \sum_{j = 1}^N \left[ i p_j \left( \frac{y_j - y_{j-1}}{\Delta t} - 2k + \gamma y_{j-1} \right) + \frac{\gamma}{2} {p_j}^2  y_{j-1} \right] \Delta t}  \ .
\end{equation}

\subsection{Computing the $y_j$ integrals}

Consider the integral over $y_j$ for some $j = 1, ..., N-1$. $y_j$ appears in the $j$th and $(j+1)$th term of our sum, so the corresponding part of the action looks like
\begin{equation}
\begin{split}
& - y_j \left\{ \frac{i p_j}{\Delta t} - \frac{i p_{j+1}}{\Delta t} + i \gamma p_{j+1} + \frac{\gamma}{2} {p_{j+1}}^2   \right\} \Delta t \\
=& - i y_j \left\{ \frac{p_j}{\Delta t} - \left[ \frac{p_{j+1}}{\Delta t} - \gamma p_{j+1} + i\frac{\gamma}{2} {p_{j+1}}^2 \right]  \right\} \Delta t \\
=& - i y_j \left\{ p_j - \left[ \left( 1 - \gamma \Delta t \right) p_{j+1} + i\frac{\gamma \Delta t}{2} {p_{j+1}}^2 \right]  \right\} \ .
\end{split}
\end{equation}
Of course, these integrals are all trivial, since the $y_j$ variables are decoupled from each other. We have
\begin{equation}
\int_0^{\infty} dy_j \ e^{- i y_j \left\{ p_j - \left[ \left( 1 - \gamma \Delta t \right) p_{j+1} + i\frac{\gamma \Delta t}{2} {p_{j+1}}^2 \right]  \right\}} = \frac{1}{i} \frac{1}{p_j - \left[ \left( 1 - \gamma \Delta t \right) p_{j+1} + i\frac{\gamma \Delta t}{2} {p_{j+1}}^2 \right]  }
\end{equation}
for $j = 1, ..., N-1$, leaving no more $y_j$ variables (except $y_0$ and $y_N$, which are constants).  The first $N-1$ momentum variable integrals, it turns out, are easily done contour integrals. Schematically, they look like
\begin{equation}
\frac{1}{2\pi i} \int_{-\infty}^{\infty} \ \frac{\exp\left[ f(p_j) \right]}{p_j - \left[ \left( 1 - \gamma \Delta t \right) p_{j+1} + i\frac{\gamma \Delta t}{2} {p_{j+1}}^2 \right]} \ dp_j
\end{equation}
where $f(p_j)$ is a stand-in for whatever the $p_j$-dependence of the remaining action is. Using Cauchy's integral formula \cite{saff2003, lang2013}, we can easily evaluate this integral as
\begin{equation}
\frac{1}{2\pi i} \int_{-\infty}^{\infty} \ \frac{\exp\left[ f(p_j) \right]}{p_j - \left[ \left( 1 - \gamma \Delta t \right) p_{j+1} + i\frac{\gamma \Delta t}{2} {p_{j+1}}^2 \right]} \ dp_j = \exp\left[ f\left( \left( 1 - \gamma \Delta t \right) p_{j+1} + i\frac{\gamma \Delta t}{2} {p_{j+1}}^2 \right) \right] \ .
\end{equation}
This means that the net effect of doing the first $N-1$ momentum variable integrals is to implement a specific constraint on the relationship between the $p_j$ (for $j = 1, ..., N-1$) and $p_N$. The constraint is that
\begin{equation} \label{eq:precurrence}
p_j = (1 - c) p_{j+1} + i \frac{c}{2} {p_{j+1}}^2
\end{equation}
for $j = 1, ..., N-1$, where we have defined $c := \gamma \Delta t$ to ease notation. 

\subsection{Approximately solving the $p_j$ recurrence relation}

We would like to solve Eq. \ref{eq:precurrence} in closed form, so that we can write any $p_j$ in terms of $p_N$, and then do the integral over $p_N$ to complete the calculation. However, this is probably not possible---quadratic recurrence relations like these (so-called `quadratic maps') are only known to be exactly solvable in specific special cases \cite{qmapmathworld, aho1973}.

Instead, we will \textit{approximately} solve this recurrence relation, by noting that it can be rewritten as
\begin{equation}
p_j = p_{j+1} + \left[ - \gamma p_{j+1} + i \frac{\gamma}{2} p_{j+1}^2 \right] \Delta t \ .
\end{equation}
How does this help? This looks like an Euler time step for the ordinary differential equation
\begin{equation} \label{eq:papproxDE}
\dot{p} = - \gamma p + i \frac{\gamma}{2} p^2 \ ,
\end{equation}
and it is exactly that in the $\Delta t \to 0$ limit. While Eq. \ref{eq:precurrence} can't be solved exactly, Eq. \ref{eq:papproxDE} can. The solution is
\begin{equation}
p(t) = - i + \tan\left[ C + i \frac{\gamma}{2} t \right] \ ,
\end{equation}
where $C$ is fixed by our initial condition. Since we `start' the recurrence at $p_N$ (and with each time step decrease the index by one; the reverse ordering of the indices is confusing here), we have
\begin{equation}
p_N + i = \tan C \ \implies \ C = \tan^{-1}(i + p_N) \ ,
\end{equation}
so
\begin{equation}
p(t) = - i + \tan\left[ \tan^{-1}(i + p_N)  + i \frac{\gamma}{2} t \right] \ .
\end{equation}
For our purposes, this means
\begin{equation}
p_j \approx - i + \tan\left[ \tan^{-1}(i + p_N)  + i \frac{\gamma}{2} (N-j) \Delta t \right] 
\end{equation}
for $j = 1, ..., N$. 

\subsection{Simplifying the remainder of the action}

Now that we have approximately solved Eq. \ref{eq:precurrence}, what is left of the action? We have
\begin{equation} \label{eq:whatsleft}
- S_{rest} = i y_0 \left[ (1 - c) p_1 + i \frac{c}{2} p_1^2 \right] - i p_N y_N + 2 k i \Delta t \sum_{j = 1}^N p_j \ .
\end{equation}
Note that
\begin{equation}
(1 - c) p_1 + i \frac{c}{2} p_1^2 = p_{1} + \left[ - \gamma p_{1} + i \frac{\gamma}{2} p_{1}^2 \right] \Delta t \ ,
\end{equation}
i.e. it corresponds to taking another Euler time step. Hence,
\begin{equation} \label{eq:firstpart}
\begin{split}
(1 - c) p_1 + i \frac{c}{2} p_1^2 &\approx - i + \tan\left[ \tan^{-1}(i + p_N)  + i \frac{\gamma}{2} N \Delta t \right] \\
&= - i + \tan\left[ \tan^{-1}(i + p_N)  + i \frac{\gamma}{2} T \right] \ .
\end{split}
\end{equation}
Meanwhile, the sum in Eq. \ref{eq:whatsleft} can be rewritten as a Riemann sum:
\begin{equation}
\begin{split}
2 k i \Delta t \sum_{j = 1}^N p_j  &= 2 k i \Delta t \sum_{j = 1}^N - i + \tan\left[ \tan^{-1}(i + p_N)  + i \frac{\gamma}{2} (N-j) \Delta t \right]   \\
&= 2 k T + 2 k i \Delta t \sum_{j = 1}^N \tan\left[ \tan^{-1}(i + p_N)  + i \frac{\gamma}{2} (N-j) \Delta t \right]   \\
&= 2 k T + 2 k i \Delta t \sum_{j = 0}^{N-1} \tan\left[ a  + i \frac{\gamma}{2} j \Delta t \right]   \\
&= 2 k T + 4 \mu  \sum_{j = 0}^{N-1} \tan\left[ a  + i \frac{\gamma}{2} j \Delta t \right]  i \frac{\gamma}{2} \Delta t \\
&= 2 k T + 4 \mu  \sum_{j = 0}^{N-1} \tan\left[ a  + j \Delta x \right]  \Delta x 
\end{split}
\end{equation}
where $a := \tan^{-1}(i + p_N)$ and $\Delta x := i \frac{\gamma}{2} \Delta t$. Approximating it as an integral, we have
\begin{equation} \label{eq:secondpart}
\begin{split}
& 2 k T + 4 \mu  \sum_{j = 0}^{N-1} \tan\left[ a  + j \Delta x \right]  \Delta x \\
&\approx 2 k T + 4 \mu  \int_a^{a + i \frac{\gamma}{2} T} \tan x  \ dx \\
&= 2 k T - 4 \mu  \left. \log\left( \cos x \right)  \right|_a^{a + i \frac{\gamma}{2} T} \\
&= 2 k T - 4 \mu   \log\left[ \frac{\cos\left( a + i \frac{\gamma}{2} T \right)}{\cos a} \right]   \ .
\end{split}
\end{equation}
Using Eq. \ref{eq:firstpart} and Eq. \ref{eq:secondpart}, we are left with the integral
\begin{equation} \label{eq:pNintegral}
e^{y_0 + 2 k T} \frac{1}{2\pi} \int_{-\infty}^{\infty} \ \exp\left\{i y_0 \tan\left[ \tan^{-1}(i + p_N)  + i \frac{\gamma}{2} T \right] - i p_N y_N - 4 \mu   \log\left[ \frac{\cos\left( a + i \frac{\gamma}{2} T \right)}{\cos a} \right] \right\} \ dp_N 
\end{equation}
just over $p_N$. Does this answer make sense? One sanity check is to make sure that it reduces to a delta function in the $T \to 0$ limit. Taking $T \to 0$ yields
\begin{equation}
\begin{split}
& e^{y_0 } \frac{1}{2\pi} \int_{-\infty}^{\infty} \ \exp\left\{i y_0 \tan\left[ \tan^{-1}(i + p_N)   \right] - i p_N y_N - 4 \mu   \log\left[ \frac{\cos a}{\cos a} \right] \right\} \ dp_N \\
=& e^{y_0 } \frac{1}{2\pi} \int_{-\infty}^{\infty} \ \exp\left\{ -y_0 + i y_0 p_N  - i p_N y_N  \right\} \ dp_N \\
=& \frac{1}{2\pi} \int_{-\infty}^{\infty} \ \exp\left\{ i p_N (y_0 - y_N)   \right\} \ dp_N \\
=& \delta(y_0 - y_N) \\
=& \delta(x_0 - x_N)
\end{split}
\end{equation}
as expected.

\subsection{Simplifying the tangent term}
\label{sec:tangent}

Let's parse the overall integral in Eq. \ref{eq:pNintegral} piece by piece. Note that
\begin{equation}
\begin{split}
\tan\left[ \tan^{-1}(i + p_N)  + i \frac{\gamma}{2} T \right] &= \frac{i + p_N + \tan\left( i \frac{\gamma}{2} T \right)}{1 - (i + p_N)\tan\left( i \frac{\gamma}{2} T \right)}  \\
&= \frac{i + p_N + i \tanh\left( \frac{\gamma}{2} T \right)}{1 - (i + p_N) i \tanh\left( \frac{\gamma}{2} T \right)}  \\
&= \frac{\cosh\left( \frac{\gamma}{2} T \right)(i + p_N) + i \sinh\left( \frac{\gamma}{2} T \right)}{\cosh\left( \frac{\gamma}{2} T \right) - (i + p_N) i \sinh\left( \frac{\gamma}{2} T \right)}  \ .
\end{split}
\end{equation}
Since $\sinh\left( \frac{\gamma}{2} T \right) + \cosh\left( \frac{\gamma}{2} T \right) = e^{\frac{\gamma}{2} T}$, this becomes
\begin{equation}
\begin{split}
& \frac{i e^{\frac{\gamma}{2} T} + p_N \cosh\left( \frac{\gamma}{2} T \right)}{e^{\frac{\gamma}{2} T} - i p_N \sinh\left( \frac{\gamma}{2} T \right)}  \\
=& \frac{i + \frac{p_N}{2} \left[ 1 + e^{-\gamma T} \right]}{1 - i \frac{p_N}{2} \left[ 1 - e^{- \gamma T} \right]}  \\
=& \frac{i - \frac{p_N}{2} \left[ 1 - e^{-\gamma T} \right] +  \frac{p_N}{2} \left[ 1 + e^{-\gamma T} \right] + i \frac{p_N^2}{4} \left[ 1 - e^{- 2\gamma T} \right]}{1 + \frac{p_N^2}{4} \left[ 1 - e^{- \gamma T} \right]^2}  \\
=& \frac{i + p_N e^{-\gamma T} + i \frac{p_N^2}{4} \left[ 1 - e^{- 2\gamma T} \right]}{1 + \frac{p_N^2}{4} \left[ 1 - e^{- \gamma T} \right]^2}  \ .
\end{split}
\end{equation}
Also note that
\begin{equation}
\begin{split}
i \frac{p_N^2}{4} \left[ 1 - e^{- 2\gamma T} \right] &= i \left( \frac{1 + e^{- \gamma T}}{1 - e^{- \gamma T}} \right) \frac{p_N^2}{4} \left( 1 - e^{- \gamma T} \right)^2 \\
&= i \left( \frac{1 + e^{- \gamma T}}{1 - e^{- \gamma T}} \right) \left[ 1 + \frac{p_N^2}{4} \left( 1 - e^{- \gamma T} \right)^2 \right] - i \left( \frac{1 + e^{- \gamma T}}{1 - e^{- \gamma T}} \right) \ .
\end{split}
\end{equation}
We now have
\begin{equation}
\begin{split}
\tan\left[ \tan^{-1}(i + p_N)  + i \frac{\gamma}{2} T \right] &= \frac{i + p_N e^{-\gamma T} + i \frac{p_N^2}{4} \left[ 1 - e^{- 2\gamma T} \right]}{1 + \frac{p_N^2}{4} \left[ 1 - e^{- \gamma T} \right]^2} \\
&= i \left( \frac{1 + e^{- \gamma T}}{1 - e^{- \gamma T}} \right) + \frac{i + p_N e^{-\gamma T} - i \left( \frac{1 + e^{- \gamma T}}{1 - e^{- \gamma T}} \right)}{1 + \frac{p_N^2}{4} \left[ 1 - e^{- \gamma T} \right]^2} \\
&= i \left( \frac{1 + e^{- \gamma T}}{1 - e^{- \gamma T}} \right) + \frac{p_N e^{-\gamma T} - 2i \left( \frac{e^{- \gamma T}}{1 - e^{- \gamma T}} \right)}{1 + \frac{p_N^2}{4} \left[ 1 - e^{- \gamma T} \right]^2} \\
&= i \left( \frac{1 + e^{- \gamma T}}{1 - e^{- \gamma T}} \right) + \frac{2 e^{-\gamma T}}{1 - e^{- \gamma T}} \frac{ \left[ \frac{p_N}{2}(1 - e^{- \gamma T}) - i \right]}{1 + \frac{p_N^2}{4} \left[ 1 - e^{- \gamma T} \right]^2} \\
&= i \left( \frac{1 + e^{- \gamma T}}{1 - e^{- \gamma T}} \right) + \frac{2 e^{-\gamma T}}{1 - e^{- \gamma T}} \frac{ 1}{\frac{p_N}{2}(1 - e^{- \gamma T}) + i} \ .
\end{split}
\end{equation}

\subsection{Simplifying the log cosine term}
\label{sec:logcosine}

We are concerned with the term
\begin{equation}
- 4 \mu   \log\left[ \frac{\cos\left( a + i \frac{\gamma}{2} T \right)}{\cos a} \right] 
\end{equation}
in the exponential of Eq. \ref{eq:pNintegral}. Note,
\begin{equation}
\begin{split}
\frac{\cos\left( a + i \frac{\gamma}{2} T \right)}{\cos a} &= e^{\frac{\gamma}{2} T} - i \sinh\left( \frac{\gamma}{2} T \right) p_N \\
&= e^{\frac{\gamma}{2} T} \left[ 1 - i (1 - e^{- \gamma T}) \frac{p_N}{2} \right] \ .
\end{split}
\end{equation}

\subsection{Finishing the calculation}

Using the results of Sec. \ref{sec:tangent} and \ref{sec:logcosine}, the integral in Eq. \ref{eq:pNintegral} (which we will denote by $I$) can be rewritten as
\begin{equation}
\begin{split}
I =& e^{y_0 + 2 k T} \frac{1}{2\pi} \int_{-\infty}^{\infty} \ \exp\left\{i y_0 \tan\left[ \tan^{-1}(i + p_N)  + i \frac{\gamma}{2} T \right] - i p_N y_N - 4 \mu   \log\left[ \frac{\cos\left( a + i \frac{\gamma}{2} T \right)}{\cos a} \right] \right\} \ dp_N \\
=& e^{y_0 + 2 k T} \frac{1}{2\pi} \int_{-\infty}^{\infty} \ \frac{\exp\left\{i y_0 \left[ i \left( \frac{1 + e^{- \gamma T}}{1 - e^{- \gamma T}} \right) + \frac{2 e^{-\gamma T}}{1 - e^{- \gamma T}} \frac{ 1}{\frac{p_N}{2}(1 - e^{- \gamma T}) + i}  \right] - i p_N y_N  - 2 k T \right\}}{\left[ 1 - i (1 - e^{- \gamma T}) \frac{p_N}{2} \right]^{4\mu}} \ dp_N \\
=& e^{- \frac{2 y_0 e^{- \gamma T}}{1 - e^{- \gamma T}}} \frac{1}{2\pi} \int_{-\infty}^{\infty} \ \frac{\exp\left\{i y_0 \left[ \frac{2 e^{-\gamma T}}{1 - e^{- \gamma T}} \frac{ 1}{\frac{p_N}{2}(1 - e^{- \gamma T}) + i}  \right] - i p_N y_N  \right\}}{\left[ 1 - i (1 - e^{- \gamma T}) \frac{p_N}{2} \right]^{4\mu}} \ dp_N \ .
\end{split}
\end{equation}
Make a change of variables
\begin{equation}
\begin{split}
z &:= \frac{p_N}{2} (1 - e^{- \gamma T}) \\
\frac{2}{1 - e^{- \gamma T}} dz &= dp_N
\end{split}
\end{equation}
so we have
\begin{equation}
\begin{split}
I =& e^{- \frac{2 y_0 e^{- \gamma T}}{1 - e^{- \gamma T}}} \frac{1}{\pi (1 - e^{-\gamma T})} \int_{-\infty}^{\infty} \ \frac{\exp\left\{\frac{2 i y_0 e^{-\gamma T}}{1 - e^{- \gamma T}} \frac{1}{z + i}   - \frac{2 i y_N}{1 - e^{- \gamma T}} z  \right\}}{\left[ 1 - i z \right]^{4\mu}} \ dz \\
=& e^{- \frac{2 y_0 e^{- \gamma T}}{1 - e^{- \gamma T}}} \frac{1}{\pi (1 - e^{-\gamma T})} \int_{-\infty}^{\infty} \ \frac{\exp\left\{\frac{2 y_0 e^{-\gamma T}}{1 - e^{- \gamma T}} \frac{1}{1 - iz}   - \frac{2 i y_N}{1 - e^{- \gamma T}} z  \right\}}{\left[ 1 - i z \right]^{4\mu}} \ dz \ .
\end{split}
\end{equation}
For convenience, define
\begin{equation}
\begin{split}
A &:= \frac{2 y_N}{1 - e^{- \gamma T}} \\
B &:= \frac{2 y_0 e^{-\gamma T}}{1 - e^{- \gamma T}} 
\end{split}
\end{equation}
so that our integral reads
\begin{equation}
I = e^{- \frac{2 y_0 e^{- \gamma T}}{1 - e^{- \gamma T}}} \frac{1}{\pi (1 - e^{-\gamma T})} \int_{-\infty}^{\infty} \ \frac{\exp\left\{B \frac{1}{1 - iz}   - i A z  \right\}}{\left[ 1 - i z \right]^{4\mu}} \ dz \ .
\end{equation}
Now we will Taylor expand the integrand so we have
\begin{equation}
I =  \frac{e^{- \frac{2 y_0 e^{- \gamma T}}{1 - e^{- \gamma T}}}}{\pi (1 - e^{-\gamma T})} \sum_{k = 0}^{\infty} \frac{B^k}{k!} \int_{-\infty}^{\infty} \ \left( 1 - i z \right)^{-(4\mu + k)} \exp\left\{- i A z  \right\} \ dz \ .
\end{equation}
At this point we will use a special result from a table of integrals. The specific result is from Gradshteyn and Ryzhik \cite{gradshteyn2014} (ET I 118(3), in section 3.382, on pg. 365), and says that
\begin{equation}
\int_{- \infty}^{\infty} (\beta - i x)^{- \nu} e^{- i p x} dx = \frac{2 \pi p^{\nu - 1} e^{- \beta p}}{\Gamma(\nu)}
\end{equation}
for $p > 0$, $\text{Re}(\nu) > 0$, and $\text{Re}(\beta) > 0$. Using it,
\begin{equation}
\begin{split}
I =&  \frac{e^{- \frac{2 y_0 e^{- \gamma T}}{1 - e^{- \gamma T}}}}{\pi (1 - e^{-\gamma T})} \sum_{k = 0}^{\infty} \frac{B^k}{k!} \left[ \frac{2 \pi A^{4\mu + k - 1} e^{- A}}{\Gamma(4\mu + k)} \right] \\
=&  2 \frac{e^{- \frac{2 y_0 e^{- \gamma T}}{1 - e^{- \gamma T}}}}{(1 - e^{-\gamma T})} A^{4\mu - 1} e^{-A} \sum_{k = 0}^{\infty} \frac{(A B)^k}{k! \Gamma(4\mu + k)}  \\
=&  2 \frac{e^{- \frac{2 y_0 e^{- \gamma T}}{1 - e^{- \gamma T}}}}{(1 - e^{-\gamma T})} A^{4\mu - 1} e^{-A} \frac{(\sqrt{AB})^{4 \mu - 1}}{(\sqrt{AB})^{4 \mu - 1}} \sum_{k = 0}^{\infty} \frac{(A B)^k}{k! \Gamma(4\mu + k)}  \ .
\end{split}
\end{equation}
Since
\begin{equation}
I_{\nu}(z) = \left( \frac{z}{2} \right)^{\nu} \sum_{k = 0}^{\infty} \frac{\left( \frac{z^2}{4} \right)^k}{k! \Gamma(\nu + k)} \ ,
\end{equation}
we can write our result in terms of a modified Bessel function $I_{4\mu - 1}$:
\begin{equation}
I =  2 \frac{e^{- \frac{2 y_0 e^{- \gamma T}}{1 - e^{- \gamma T}}}}{(1 - e^{-\gamma T})} A^{4\mu - 1} e^{-A} \frac{1}{(\sqrt{AB})^{4 \mu - 1}} I_{4\mu - 1}\left( 2 \sqrt{AB} \right)  \ .
\end{equation}
Modulo some rewriting, this is actually the final result for the transition probability. Note,
\begin{equation}
\begin{split}
\frac{A^{4\mu - 1}}{(\sqrt{AB})^{4 \mu - 1}} &= \frac{2^{4\mu - 1} y^{4\mu - 1}}{(1 - e^{- \gamma T})^{4 \mu - 1}} \frac{(1 - e^{- \gamma T})^{4 \mu - 1}}{(2 y 2 y_0 e^{- \gamma T})^{(4\mu - 1)/2}} = \frac{2^{4\mu - 1} y^{4\mu - 1}}{(2 y 2 y_0 e^{- \gamma T})^{(4\mu - 1)/2}} 
\end{split}
\end{equation}
and
\begin{equation}
\begin{split}
\exp{- \frac{2 y_0 e^{- \gamma T}}{1 - e^{- \gamma T}} - A} &= \exp{- \frac{2 y_0 e^{- \gamma T}}{1 - e^{- \gamma T}} - \frac{2 y}{1 - e^{- \gamma T}} } \\
&= \exp{- \frac{2 y_0 e^{- \gamma T}}{1 - e^{- \gamma T}} + \frac{-2 y (1 - e^{- \gamma T}) - 2 y e^{- \gamma T}}{1 - e^{- \gamma T}} } \\
&= \exp{-2 y - \frac{2 (y_0 + y) e^{- \gamma T}}{1 - e^{- \gamma T}} } \ .
\end{split}
\end{equation}
Finally, we have
\begin{equation}
\begin{split}
P(x, t; x_0, t_0) &= 2 e^{-2y} \frac{1}{(1 - e^{-\gamma T})}  \frac{2^{4\mu - 1} y^{4\mu - 1}}{(2 y 2 y_0 e^{- \gamma T})^{(4\mu - 1)/2}}  e^{- \frac{2 (y_0 + y) e^{- \gamma T}}{1 - e^{- \gamma T}} } I_{4\mu - 1}\left( 2 \sqrt{AB} \right)  \\
&= 2^{4\mu} (x + \mu)^{4\mu - 1} e^{-2(x + \mu)}  \frac{1}{(2 y 2 y_0 e^{- \gamma T})^{(4\mu - 1)/2} (1 - e^{-\gamma T})}  e^{- \frac{2 (y_0 + y) e^{- \gamma T}}{1 - e^{- \gamma T}} } I_{4\mu - 1}\left( \frac{2 \sqrt{2 y 2 y_0 e^{- \gamma T}}}{1 - e^{- \gamma T}} \right)  \ .
\end{split}
\end{equation}
Since $w_0 = 2 y_0$ and $w = 2 y$, this is the same as the result we derived earlier by the method of eigenfunction expansion (Eq. \ref{eq:ptrans}).

\section{Discussion}
\label{sec:discussion}

We exactly computed the MSRJD path integral (Eq. \ref{eq:MSRJD_gill}) \cite{vastola2019} corresponding to the chemical birth-death process with Gillespie noise (Eq. \ref{eq:CLE}), without recourse to perturbative or asymptotic expansions. We also showed that the result agrees with what one would find using the well-known method of eigenfunction expansion. 

Because its use yields the correct result in this nontrivial case, we have increased confidence that the stochastic path integral derived in \cite{vastola2019} is correct. We imagine that this problem could become a benchmark test for different stochastic path integral approaches, since it is exactly solvable but has state-dependent noise; it would be interesting to see if alternative approaches reproduce the correct result.

Although the calculation in Sec. \ref{sec:pathintcalc} was lengthy, it was straightforward, and did not involve much sophisticated mathematics. While it is clear from comparing the lengths of Sec. \ref{sec:eigexpansion} and Sec. \ref{sec:pathintcalc} that path integration is not the preferred way to solve this particular problem, it is possible that for more complicated problems an approximate path integral approach will yield nontrivial insights even when the Fokker-Planck equation seems intractable.

Few path integrals are exactly solvable, and exactly solving the path integrals of even textbook problems (e.g. the particle in a box \cite{janke1979}, the harmonic oscillator \cite{feynman2010}, and the hydrogen atom \cite{kleinert2009}) can involve considerable technical challenges. It is interesting to note that each of these three paradigmatic quantum problems has a precise stochastic dynamics analogue: unbiased additive noise diffusion on a finite domain is like the particle in a box; the chemical birth-death process with additive noise (which was solved using path integrals and other methods in \cite{vastolaADD2019}) is like the harmonic oscillator; and the problem studied in this paper is like the hydrogen atom. Indeed, the solution to the hydrogen atom also involves associated Laguerre polynomials in its eigenfunctions \cite{griffiths2018}, and a modified Bessel function of the first kind in its propagator \cite{duru1982, kleinert2009}. 

Interestingly, although the MSRJD path integral approach to this problem is straightforward, it is not even clear if the Onsager-Machlup path integral is well-defined, since it yields many integrals of the form $e^{-1/x}$. This problem seems to be analogous to well-known problems with the hydrogen atom's configuration space path integral (see pg. 934 of Kleinert \cite{kleinert2009}) and partition function \cite{lucena1995, blinder1995, basu1999}.

The model studied in this paper is strikingly similar to the Cox-Ingersoll-Ross (CIR) model, a model used in mathematical finance to describe the time evolution of interest rates \cite{cir1985, brown1986, hull2003}. In principle, the MSRJD path integral method presented here can reproduce calculations like \cite{bennati1999, lemmens2008} for the CIR process and similar models from finance. 

\section{Conclusion}

In this paper, we exactly calculated the MSRJD path integral corresponding to the chemical birth-death process with Gillespie noise (a canonical toy problem from chemical kinetics), and verified our result using an eigenfunction expansion solution of the Fokker-Planck equation. Our result suggests that the stochastic path integrals from \cite{vastola2019} are valid even in the case of state-dependent noise, and can be confidently applied to other problems in continuous stochastic dynamics.

\section{Acknowledgments}

This work was supported by NSF Grant \# DMS 1562078.

%%%%%%%%%%

\end{document}